\documentclass{article}

\usepackage{arxiv}
\usepackage{tabularx}
\usepackage{multirow}
\usepackage{multicol}
\usepackage[utf8]{inputenc} 
\usepackage[T1]{fontenc}    
\usepackage{hyperref}       
\usepackage{url}            
\usepackage{csquotes}
\usepackage{booktabs}       
\usepackage{amsfonts}       
\usepackage{nicefrac}       
\usepackage{microtype}      
\usepackage{lipsum}
\usepackage{fancyhdr}       
\usepackage{graphicx}       
\graphicspath{{media/}}     
\usepackage{hyperref}
\title{Responsible AI in Open Ecosystems: Reconciling Innovation with Risk Assessment and Disclosure}
\usepackage{geometry}
\usepackage{longtable}
\usepackage{tabularx, booktabs}
\usepackage{float}
\newcolumntype{s}{>{\hsize=.5\hsize}X}

\author{Mahasweta Chakraborti\\
  University of California, Davis, CA \\
\texttt{mchakraborti@ucdavis.edu}
\And
Bert Joseph Prestoza\\
  University of California, Davis, CA \\
\texttt{bertjosephprestoza@gmail.com}
\And
Nicholas Vincent\\
  Simon Fraser University, British Columbia\\
\texttt{nicholas\_vincent@sfu.ca}
\And
Seth Frey\\
  University of California, Davis, CA \\
\texttt{sethfrey@ucdavis.edu}}

\begin{document}
\maketitle

\begin{abstract}
The rapid scaling of AI has spurred a growing emphasis on ethical considerations in both development and practice. This has led to the formulation of increasingly sophisticated model auditing and reporting requirements, as well as governance frameworks to mitigate potential risks to individuals and society. At this critical juncture, we review the practical challenges of promoting responsible AI and transparency in informal sectors like OSS that support vital infrastructure and see widespread use.  We focus on how model performance evaluation may inform or inhibit probing of model limitations, biases, and other risks. Our controlled analysis of 7903 Hugging Face projects found that risk documentation is strongly associated with evaluation practices. Yet, submissions (N=789) from the platform's most popular competitive leaderboard showed less accountability among high performers. Our findings can inform AI providers and legal scholars in designing interventions and policies that preserve open-source innovation while incentivizing ethical uptake.

\end{abstract}


\section{Introduction}


In recent years, we have witnessed the adoption of artificial intelligence (AI) across various individual, collective, and public enterprises, including education, business, and research. This widespread implementation has been argued to boost productivity, enhance manufacturing, accelerate development, and facilitate the provisioning of critical services and infrastructure at unprecedented levels and reach (see e.g. Chapter 4 of the AI Index Report for an overview of specific claims along these lines \cite{Maslej2024ArtificialII}). The scalability offered by these technologies may revolutionize numerous industries and promote further investment in AI innovation for improved service delivery across diverse domains.

Nascent technologies often face skepticism and scrutiny before earning public trust ~\cite{floridi_ai4peopleethical_2018}. Therefore, potential risks in AI need to be recognized and addressed at every stage of development, such as representative quality of training data~\cite{hutchinson_towards_2021}, algorithmic designs~\cite{dwork_fairness_2012}, and learning objectives \cite{weidinger_ethical_2021}. AI artifacts are also highly prone to inappropriate uses \cite{contractor2022behavioral,mcduff2024standardization}. As AI's market impact and reach in people's lives continues to expand, there have also been corresponding calls for ethical training and regulatory oversight for providers and deployers.

Model evaluation is a standard component of the AI development and deployment cycle. In its most common form, evaluation involves testing models on held-out datasets to understand how well a model has learned. Such benchmarking is essential for assessing novelty against the state-of-the-art, deciding whether the model is suitable for widespread use, improving training, and informing design choices for further innovation. Today, competitive benchmarking against other models is a popular form of evaluation, and high performers garner legitimacy from users and investors, enjoy market visibility, and even steer development and consumption ~\cite{dehghani_benchmark_2021, ethayarajh-jurafsky-2020-utility,raji_ai_2021}.

With rising stakes, responsible developers are increasingly expected to use evaluations not only to assess model capabilities but also to \textit{recognize its limitations}. Holistic evaluation goes beyond simply reporting gross accuracy and encourages probing edge cases, measuring biases in predictions for specific domains and vulnerable subpopulations, and cautions against failure modes ~\cite{liang_holistic_2023,bommasani_opportunities_2022}.  Depending upon the criticality of the application (e.g., medical diagnostics or defense), evaluation may demand nuanced expertise, where the developer needs to invest in understanding methodologies, conducting different tests, and selecting appropriate metrics. Empirical testing can provide developers, managers, and investors with necessary information regarding broader applicability and sociotechnical impact, thereby mitigating liability and harm. Importantly, these assessments should be documented and reported in an interpretable, accessible form that can empower experts and ordinary users to make informed choices over model use. 

Our work explores current evaluation practices and developer accountability in informal sectors. Open-source AI is a rising player with a considerable market presence and increasing corporate adoption. Prized for fast-track innovation through crowd-sourced contributions, OSS projects have garnered significant attention from researchers in online communities and collaborative work. Yet open-source ecosystems face unique challenges in actuating responsible development; simply put, the governance and power structures tend to be more decentralized than in corporations, universities, or other institutions that might develop AI systems. Fostering accountability and standardization would necessitate collaborative monitoring of risk documentation, inadequate evaluations or malpractice ~\cite{venturebeatOpenSource,ethayarajh-jurafsky-2020-utility,balloccu2024leak}, and downstream misuse. Their largely informal, decentralized organization might complicate agreement over testing requirements, documentation standards, and monitoring of their uptake.  Holistic evaluations ~\cite{liang_holistic_2023} may also bring additional costs and effort for small communities. Lastly, communities may be apprehensive about retaining their user base with stringent risk disclosure requirements.

We conduct an in-depth dataset documentation study of metadata on Hugging Face (HF), one of the largest open-source AI hosting platforms. In particular, we focus on collecting data and conducting quantitative analyses on the evaluation and risk documentation practices among projects and model the relationship between the two. Our contributions include the following:

\begin{itemize}
    \item We thoroughly review documentation practices (Model cards), project organization, information management, compliance checks, and other platform support on Hugging Face. 
    \item We explore the evolution of open source AI development through the scaling of model training, applications, contribution rates, model re-use patterns, and documentation practices among models across different organizational entities 
    \item Among usable, service-ready projects, around 15.9\% and 2.2\% models contained evaluation and risk-related documentation from the developer. 
    \item In a quantitative analysis of 7903 models with controls, we found a strong, positive association between the practices of model evaluation and risk documentation practices
    \item Among 789 projects participating in HF's Open LLM leaderboard~\cite{open-llm-leaderboard-v1}, higher-performing models were found to be less likely to provide documentation on risks and limitations.
\end{itemize}

\section{Related Work}
Here, we discuss background on Open Source AI and key areas of research that motivated our dataset documentation effort and accompanying quantitative study.

\subsection{Ethics of AI, Tech Regulation, and Open Development}
\label{sec:law}

Fostering and standardizing ethical development is a nuanced challenge in open source, which, contrary to regulated for-profits and commercial entities, is historically informal and free-forming ~\cite{0ed03cec-f1bb-3949-a3ba-f465bb395084, crowston2012free,coleman2013coding}. Developer motivations, ranging from altruism to popularity, may lead to varying governance structures ~\cite{li2021code,chakraborti2024we,10.1145/3555129}. Open sourcing has led to remarkable progress in science and technology, and notable AI projects started out in the same to inspire and support collective innovation. At the same time, there have been multiple instances of biased or improperly curated training data and artifacts that can compromise regular applications  ~\cite{birhane2021multimodal, NEURIPS2023_42f22550,lee_deepfakes_2024}. Further, open source projects have also been appropriated for nefarious uses ~\cite{justiceArrestedProducing,incidentdatabaseArtificialIntelligence,marchal_generative_2024,fortuneAIChatbot}. We review some milestones in ethical discussions and regulatory approaches relevant to open-source AI.

With the growing potential of AI uses and misuse, researchers and ethicists came together early on to specify and advocate documentation guidelines for every link in the AI development pipeline. Model cards,~\cite{mitchell_model_2019,crisan_interactive_2022,arnold_factsheets_2019}, data sheets~\cite{gebru_datasheets_2018,bender_data_2018}, and other factsheets are crucial sociotechnical governance tools that keep stakeholders informed and define the scope of AI consumer applications, ultimately contributing to more transparent and accountable development practices. Responsible documentation outlines permissible use cases (possibly extending to licensing~\cite{contractor2022behavioral}), cautioning against out-of-scope applications, and discloses other anticipated risks and practical challenges. Projects generally explain any anticipated limitations in implementing the AI solution, such as predictive biases across ethnicity or gender-profession skew in generated outputs, robustness to adversarial attacks, reliability across use cases (e.g., whether a medical diagnostic system under diagnoses rare conditions), cautioning corner cases where a model is unsuitable for use. 

Consensus over development standards and reporting requirements are evolving with expanding uses of AI, leading to attempts at governance systems \cite{arnold_factsheets_2019,floridi_ai4peopleethical_2018,olteanu_social_2019,jobin_global_2019}. Legislative measures towards AI accountability also deliberated the drawbacks of strict regulation. Despite arguments in favor and against~\cite{law_open-source_2023}, open source is primarily exempted by the EU AI act, one of the first major steps towards formalized AI governance~\cite{EU_AI_Act_2024}. Yet, given product objectives and target uses, other vulnerabilities can arise from development, such as biased training data ~\cite{hutchinson_towards_2021} (content, source, representative quality, etc.), limited robustness or generalizability~\cite{li_trustworthy_2021} (particularly for high-stake uses), and other associated design choices. In Recital 89, the act strongly endorses documentation practices such as model cards and data sheets for open source developers ''to accelerate information sharing along the AI value chain, promoting trustworthy AI systems in the Union''. Therefore, open source developers must still strive towards responsible practices, address the concerns of potential stakeholders through information sharing, and institute contingencies to address risks and harms. 

Given the contemporary state of affairs, we seek a closer understanding of open source aspirations to inform the design of guardrails that foster mindfulness and social responsibility while also preserving reasonable developer freedom. 

\subsection{Evaluation  Practices and Responsible AI}
\label{sec:respAI}
Evaluation and testing have been historically integral in informing innovation and consumer use of emergent technologies. As we work on making AI more capable, there are benefits to be seen from striving towards a 99\% accurate solution from a 98\% accurate solution. These gross accuracy metrics indeed serve to inform design decisions. However, model valuation through accuracy is only one of several aspects that concern AI use and governance~\cite{mokander_auditing_2023,falCO_governing_2021}. We review current approaches to AI evaluation, which inform how we analyze our collected data and interpret our findings.

A bid for greater accuracy through standardization led to the rise of evaluation benchmarks. As benchmarks became more widely adopted, they evolved into popular, competitive leaderboards~\cite{wang2019superglue,wang2018glue,rajpurkar2016squad,srivastava2022beyond,bojar2017findings,deng2009imagenet,lecun1998gradient}. These initiatives draw attention and participation, and have been instrumental in the rapid progress of AI by serving as a recognition-based incentive for continuous, incremental innovation. However, strides towards predictive efficiency have also culminated in a uni-dimensional emphasis on metric-races and ranking. 

Over time, several researchers have observed practical limitations with bench-marking and leaderboards, including errors and other vulnerabilities that compromise their validity or even obfuscate limitations and risks ~\cite{dehghani_benchmark_2021,gema_are_2024,bowman2021will,sainz2023nlp}. Despite their usefulness as a general estimation of model capabilities, held-out test accuracy from a benchmark is not a comprehensive measure of generalizability, adversarial robustness or risk tolerance. Therefore, the error rate on a benchmark may not always reflect the population error rate ~\cite{jia2017adversarial,zhang2020adversarial,ethayarajh-jurafsky-2020-utility,dehghani_benchmark_2021,arora2021rip,blum2015ladder} Leaderboards are often dominated by highly over-parameterized, complex, and energy-inefficient models, sometimes overfitted to benchmarks ~\cite{ethayarajh-jurafsky-2020-utility,hardt2017climbing,arora2021rip,dehghani_benchmark_2021,blum2015ladder}. Because of the visibility enjoyed by top performing developers, rampant attempts have been observed to game leaderboards, such as multiple submissions and tweaks, to maximize rank ~\cite{hardt2017climbing}. Benchmarks also generally lack transparency and fail to account for model attributes crucial to design and informing use, such as compactness, fairness, inference speed, and energy footprint, to name a few. ~\cite{ethayarajh-jurafsky-2020-utility,raji_ai_2021}. Moreover, as rankings are generally based on models' aggregate performance (e.g., accuracy/F1, etc.) over a collection of tasks and datasets, submissions often conceal racial and gender biases ~\cite{bordia2019identifying,manzini2019black,rudinger2018gender,blodgett2020language}. Researchers have also noted selective pressures arising from benchmarking with unintended consequences for innovation, such as the promotion of certain architectures and algorithms over others ~\cite{dehghani_benchmark_2021, tay2021pretrained}.

With increasing recognition of the inadequacy of current evaluation practices, scholarship in AI safety has seen a notable push towards holistic approaches with expanding definitions, objectives and approaches beyond simply predictive accuracy~\cite{liang_holistic_2023,bommasani_foundation_2023,bommasani_opportunities_2022,mehrabi_survey_2021}. These include robustness against malicious or adversarial inputs~\cite{nie2020adversarial}, explainability of the model’s decision-making and intermediate processes~\cite{10.1145/3411763.3445016,10.1145/3411764.3445188}, generalizability on out-of-sample data~\cite{bartlett2021deep}, and granular testing across demographic subgroups for biased behavior. There have also been notable strides in designing evaluations to measure model fairness ~\cite{zhao2018gender, nadeem2021stereoset,raji2019actionable}. 

With the legitimacy popular leaderboards enjoy~\cite{raji_ai_2021}, high performance may easily lead developers and users to assume the generalizability of their models and undermine the need for additional efforts to stress-test for runtime risks. This is especially dangerous and requires a reassessment of benchmarking as it is. Opportunistic overfitting can be mitigated by promoting dynamic benchmarks ~\cite{nie2020adversarial,dehghani_benchmark_2021,kiela_dynabench_2021,gehrmann_gem_2021}, that are constantly updated to accommodate temporal/distributional drifts and emerging domains, newer tasks, and capabilities. Dehghani et al. and Hardt et al. further guide benchmark design to ensure judicious assessment while mitigating opportunistic submissions ~\cite{dehghani_benchmark_2021,hardt2017climbing,blum2015ladder}. Other proposed measures include confidentiality of test/hold out sets and mitigation of data leakage~\cite{lilja2024localization,deng2024benchmark} and contaminated models ~\cite{dwork2015reusable,balloccu2024leak,magar2022data}. Multi-metric leaderboards, such as SNLI, which displays model sizes alongside accuracy ~\cite{bowman2015large,snli_leaderboard}, can guide developers and users towards efficient and sustainable choices.


\subsection{Empirical studies of open source and open weight AI}
\label{sec:repo}
With rapid interest and growth of AI in open-source contributions, researchers have been exploring the potential of data-driven research and feasibility of repository mining on AI-centric development platforms and services~\cite{jiang_peatmoss_2023, jiang_ptmtorrent_2023, ait_suitability_2023, ait_hfcommunity_2024}. Software engineers have been particularly interested in modularity and artifact reuse from pre-trained model repositories or "PTMs" ~\cite{jiang_empirical_2023, gong_what_2023, taraghi_deep_2024}, and accompanying security risks and vulnerabilities \cite{jiang_empirical_2022, kathikar_assessing_2023}. Several studies have explored model documentation~\cite{pepe_how_2024,gong_what_2023,yang_navigating_2024}. Gong et al. focus on usage documentation across multiple platforms ~\cite{gong_what_2023}, while Liang et al. analyzed different sections across Hugging Face model cards and how comprehensive documentation may improve model popularity~\cite{liang_whats_2024}. Castano et al. studied carbon footprint reporting ~\cite{castano_lessons_2024}. Hugging Face's internal study found that among all model card components, respondents found the risks sections the longest and most challenging to complete~\cite{ozoani2022modelcard}. Osborne analyzed licensing and collaboration patterns, finding positively skewered patterns in contribution, engagement, and model usage~\cite{osborne_ai_2024}.

\section{Methods}

\subsection{Research Questions}
\label{sec:rq}

After careful perusal of repositories 
spanned by our review (Sec. ~\ref{sec:repo}), we base our exploration and empirical analysis on Hugging Face, the most popular of contemporary PTM repositories ~\cite{gong_what_2023} with over 0.7 mil submissions at the time of the study. Hugging Face is increasingly appealing to AI developers, even over long-standing platforms like GitHub ~\cite{ait_suitability_2023,ait_hfcommunity_2024}. This is especially true as projects scale, requiring greater storage and computing requirements. Hugging Face is a PaaS exclusively for AI/ML development, offering tooling, storage for large artifacts, and remote servers for training, testing, and hosting apps, all under a single roof. While some contemporary model directories provide official base model releases for specific libraries and frameworks (e.g., Nvidia CUDA, Tensorflow, or Pytorch model directories) or vetted research projects (e.g., ModelZoo), HF spans models from these categories alongside vast numbers of amateur submissions, community contributions as well as public and private institutions. Therefore, it provides a large enough representative sample to observe development and model adaptation practices as is and gauge developer ethics in the wild.

\subsubsection{RQ1: How is documentation of risks, limits, and biases among projects related to model evaluation?}

Evaluation and Risks are core components of standard model cards. Hugging Face's guided annotation template~\cite{huggingfaceModelCard} encourages developers to select appropriate testing procedures and benchmarks to evaluate model performance and document the results. It also recommends that such evaluation should involve testing for potential usage limitations, vulnerabilities, and biases in the model to aid sociotechnical experts in comprehensive risk documentation. 
Therefore, proper motivation, understanding, and proficiency in evaluation are expected to inculcate cognizance of responsible development practices.  We frame the research question as follows:

$Risks~and~Biases~Documented \sim Project~Covariates + Evaluation~Reported$

\subsubsection{RQ2: How is risk documentation of projects related to their accuracy?}

We may expect developers of highly accurate models to be more proficient and well-rounded and, therefore, likelier to be able to also thoroughly probe and document risks, biases, and other limitations. Yet, we explain (see Sec. ~\ref{sec:respAI}) how current trends may undermine the validity of benchmarking or even downplay the need for holistic evaluations above and beyond accuracy. RQ2 can be modeled as follows: 

$Risks~and~Biases~Documented \sim Project~Covariates + Model~Accuracy$

We pursue RQ2 on submissions to Hugging Face's first edition of the Open LLM leaderboard, which ran from May 2023 to June 2024. It drew remarkable levels of participation across different project types. Importantly, it observed rigorous community monitoring for contamination~\cite{balloccu2024leak} and other evaluation malpractices, as well as reproducibility checks to substantiate self-reported performances, thus strengthening the validity of measurements and analysis.

\subsection{Data}
\label{sec:data}
Here, we describe some of the project-level covariates we consider in our empirical data analysis and explain their inclusion, i.e., how they motivate documentation habits and accountability among projects. Table. ~\ref{tab:apps} lists details of our multi-source data collection. 

Our review of prior studies and the HF Hub codebase and documentation revealed how the information we sought is distributed across the project landing page, repository and its metadata, index tags, and finally, the model card markdown files. HF Hub uses semantic tags to index models and facilitate search, which are often auto-detected or parsed from the YML component of model cards. To access repository records or model tags, we use the Hugging Face API. We focus our study on 700,072 repositories uploaded to Hugging Face as of 06/15/2024. We only include completely open repositories by filtering out 'gated' repositories whose file contents or commit history are private. 

\begin{table*}[!ht]
\begin{tabularx}{\textwidth}{|c|X|X|c|X|}
\hline
\textbf{Project Aspect} & \textbf{Variables} & \textbf{Description} & \hfil{\textbf{Type}} & \hfil{\textbf{Source}} \\ \hline
 
\multirow{2}{*}{\textbf{Model Features}} & Model Size & Number of Model Parameters  &  Numeric & Model Page \\ \cline{2-5}

                                & Training Resources &  Data samples used to 
                                train model  & Numeric & Training Data metadata (HF API) \\ \cline{2-5}

                                & Modalities &  Modalities served e.g. Computer Vision & Categorical    & Model Card metadata (HF API) \\ \cline{2-5}
                                
                                & Domain &  Specific fields of application model is trained for e.g. code analysis, medical applications & Categorical  & Training Data metadata (HF API) \\ \hline

\multirow{2}{*}{\textbf{Model Developer}} & Team Size  & Community Strength & Numeric     & Linked Developer Profile  \\ \cline{2-5}
                                & Total Models &  Development experience of contributor   & Numeric  &   Linked Developer Profile \\ \cline{2-5}
               
                                & Entity Type &  If contributor is a for or non profit, research projects, etc & Categorical    &  Linked Developer Profile \\  \hline

\multirow{2}{*}{\textbf{User Engagement}} & Likes & Total Likes from HF Users & Numeric & Model Page \\ \cline{2-5}
                         & Deployed Apps & Number of apps on HF using model & Numeric & Model Page\\ \hline

\multirow{2}{*}{\textbf{Developer Activity}} & Age & Repository Age in days & Numeric & Git History (HF API) \\ \cline{2-5}
                & Total Commits & Development activity on repository & Numeric & Git History (HF API) \\ \cline{2-5}
                & Pull Requests & Feature Additions and Contributions received & Numeric & Git History (HF API) \\ \cline{2-5}
               & Discussions & Community feedback and engagement with repo & Numeric & Git History (HF API) \\ \hline

\multirow{1}{*}{\textbf{\hfil Compliance}} & Performance Evaluation & Developer's evaluation objectives, protocols selected and results & Categorical & HF Model Card scanner and API \\ \cline{2-5} 
\multirow{1}{*}{\textbf{\hfil (Documentation Available)}}                                       & Risks, Limitations and Biases & Foreseeable harms, vulnerabilities and limitations & Categorical & HF Model Card scanner\\ \cline{2-5} 
                                      & ${CO_{2}}$ Emissions &  Model training footprint on environment & Categorical &  HF Model Card scanner and API  \\ \hline

 \multirow{1}{*}{\textbf{Competitive}} & Accuracy & Best aggregate results reported on the Open LLM Leaderboard  &   Numeric  & Leaderboard Archives     \\ \cline{2-5} 
 \multirow{1}{*}{\textbf{Benchmarking}}    & Attempts  & Number of leaderboard submissions for a single model  &    Numeric &  Leaderboard Archives   \\ \cline{2-5} 
 & Precision &  Precision used in testing e.g. 8 Bit, BF16 etc &     Categorical  & Leaderboard Archives \\ \hline 

\end{tabularx}
\caption{Data collection across Hugging Face: Variables with description.}
\label{tab:apps}
 \vspace*{-5pt}
\end{table*}


\textit{Project use, Developer activity, and Community engagement:} Git-based information, such as repository age and commit activity, were available for all open repositories, while usage/popularity metrics were available on every model's landing page. Controlling for time lets us account for documentation practices as a function of evolving development standards, conception of ethical practices, and regulatory oversight. We measure repository age as the time between project initiation (first commit) and data collection. For developer and community engagement around a model, we measure the total number of commits, pull requests, and all other discussions (including issues) on each repo. Developers seeking greater exposure and usage of their projects may practice better documentation \cite{gong_what_2023}. Several prior studies used API calls or downloads to measure model popularity. At the time of the study, Hugging Face only displayed model download stats for the current month. We use total model likes from users as a cumulative measure of popularity. Unlike its counterparts like GitHub, Hugging Face does not offer an option to fork repositories directly but allows porting to build applications called spaces. We use the total number of spaces spinning off a repository to measure model circulation.

\textit{Model application and usability:} Since AI auditing and regulation through documentation are particularly applicable for service-ready models and AI applications ~\cite{arnold_factsheets_2019,law_open-source_2023}, we screen out incomplete projects and dumps and test our hypothesis on especially well integrated, ready to use projects. Based on a review of HF documentation and semantic categories listed in the API, we identify service-ready models through at least one of the following: 
\begin{itemize}
    \item Model cards filled with detailed instructions, examples, and use cases: Detected using HF's Model card scanner
    \item Verifiable integration into the HF ecosystem (can be used for training, tuning, or inference) : Integrated models are tagged with training or deployment options within the  HF ecosystem, such as "endpoints\_compatible", "autotrain\_compatible" or have widgets enabled on their webpage for users to explore and interact with the model.
    \item Model page carries a "Use this model" feature for deployment through a developer-provided space or supported third-party platforms.
\end{itemize}
All in all, 456,545 projects out of 700,072 repositories fulfilled this criteria. 

HF tracks information on the modalities and tasks performed in index tags for most service-ready models. These span six major types: Natural Language Processing, Computer Vision, Audio, Reinforcement Learning, Tabular Data, and Multimodal. Note that a particular AI application may qualify under multiple categories, e.g., a prompt-driven image generator may be placed under Natural Language Processing (interpreting human queries), Computer Vision (image generation), Multimodal (operates across multiple modalities, i.e., image and text), and Reinforcement Learning (learning from human feedback). 

 \textit{Developer attributes:} We scrape the model landing pages and linked developer profiles for information vital for a controlled study. Hugging Face supports single-user accounts or team accounts called ''organizations." The growing importance and evolving sophistication of documentation benefits from multiple contributors and distributed responsibilities. Hugging Face's official documentation designates model auditing responsibilities across well-defined roles, such as the manager, the sociotechnical expert, and the developer. Further, information management may also depend on the type of developer or provider. In particular, commercial entities anticipating regulatory purview may ideally conduct more thorough risk assessments to avert potential liabilities from failures and misuse. Team pages contain community sizes and the type of entity owning the account, such as a company releasing 'freemium' models, an educational institution (university or classroom), or a non-profit. For all developers, we also include the total number of models they contributed as a measure of experience.

\textit{Model Scale:}  In the context of AI, scaling refers to enhancing learnability and performance by developing highly parameterized, data-intensive models. Between 2017 and 2022, parameterization in Google's language models grew from 110 million for BERT (base) ~\cite{devlin2019bert} to 540 Billion for PaLM~\cite{researchPathwaysLanguage}. While promising enhanced capabilities, Large Foundation ("Frontier") models have also seen increasing attention from ethicists and policy oversight bodies due to foreseeable market impact and consumer stakes ~\cite{heim_training_2024,bender_dangers_2021}. Recent proposals, particularly SB 1047 in California, explore graded requirements by model value. To inform ethical practice and test hypotheses around compliance behavior, it is necessary to account for emerging legal and social motivations from model valuation that may also influence documentation rigor.

Scaling solutions demand more storage and computing facilities. Comprehensive details on training and other expenses can often be challenging to obtain, be it from proprietary closed-source or informal settings like open-source. With providers seeking to scale models towards enhanced capabilities and higher performance, compute (hardware needed, floating point operations (FLOPs), etc.) comprises a significant share of development investments and directly depends on the model. Kaplan et al. experimentally validated a power law approximating the relationship between model performance and compute, model size (parameters), and training data size~\cite{kaplan_scaling_2020}. For a chosen level of performance, the compute budget follows from the requisite training data volume and model size. Recent work has validated variations of the law across other architectures, tasks, and learning paradigms~\cite{hoffmann2022training, alabdulmohsin2022revisiting, bahri2024explaining}. These rules of thumb are widespread today and instrumental in neural scaling. We control for model value through both size (number of parameters) and training data volume (number of samples). 

Diverse model sizes, non-uniform file nomenclature, and frameworks complicate the automated loading and parsing of model details such as size. Hugging Face does not track the model sizes of all repositories~\cite{huggingfaceSortingOption}. Safetensors~\cite{huggingfaceSafetensors} and GGUF~\cite{huggingfaceGGUF} are two popular tensor formats promoted and tracked by HF. Models and training checkpoints correctly stored in these formats display verified details on their pages, including the number of parameters~\cite{huggingfaceSafetensorsParamsprecision}. We obtained the parameter count for 140,783 models. 

Developers often withhold training data details ~\cite{hutchinson_towards_2021}. Reasons include but are not limited to, IP (particularly for proprietary freemium models) and licensing terms. Moreover, data provenance for transparency and explainability may be at odds with security by exposing the model to poisoning, privacy breaches, and other adversarial attacks ~\cite[p.~2]{li_trustworthy_2021}. We only consider models with openly released training data and associated details. This allows controlled hypothesis testing between evaluation and risk documentation trends while adjusting for factors across the development cycle. 

Model index tags contain links to training data provided by the developer, and Hugging Face provides size and other structured information on nearly all datasets it hosts. After selecting models with all their training data available on HF and screening out models directed to invalid dataset repositories, we obtained training data sizes for 17,260 models. Overall, 7093 models had complete model and training data size information. 

\textit{Model Knowledge Domain:} Knowledge Domain, in the context of ML, refers to the specific cases and tasks a model has learned to perform. Models are generally trained with data samples from their target domain, and high-stakes/critical applications may command greater developer accountability. E.g., minor diagnostic errors can significantly increase liabilities and derail the applicability of AI in medicine. HF tracks nine popular training data domains, including medical, finance, code generation, etc. 

\textit{Compliance Information:}

Developers chose appropriate tests and metrics to quantify model performance based on development objectives and target use. Our first hypothesis testing requires predicting models' risk and limitations documentation against the rate at which they evaluate model performance. Model cards on HF contain several distinct sections for technicalities such as data provenance, development specifications, performance, legal/copyright aspects, and social implications of the model's use. Based on the Annotation guide\footnotemark[1], the Evaluation section requires the model developer to specify objectives, protocols, and performance results. Ideally, these should be selected to ensure domain accuracy, demographic fairness i.e. performance specifically tested across relevant user groups, and foreseeable error contexts specific to the model's use cases. Working with the developer, the sociotechnical expert generally fills out portions titled "Bias, Risks, and Limitations". They are expected to interpret all aspects of the development, from data to evaluation results and intended uses, to explain foreseeable harms and misunderstandings (including but not limited to: the model's propensity towards discrimination and stereotyping, predictive skew across demographic subgroups, robustness, outlining and forbidding use cases beyond development goals etc) and other limitations. They may optionally include warnings and mitigation strategies.

Model cards are the default face of a model's landing page, rendered from the repository's 'README.md' file\footnotemark[1]. This file contains text and a machine-parseable YAML header. A repository lacking a "README.md" model card will create a blank landing page with no information. We measure such projects as non-compliant, lacking any documentation or evaluation. HF's society and ethics team recently developed a regulatory tool to scan the text portions to check if certain sections have been filled out. Per the official, Annotated Guideline\footnotemark[1], evaluation details and carbon emissions may be provided in the header or the text. Risk information is generally more descriptive and only reported in the main model card text. 

Using the text scanner and API, we analyze all model cards to detect whether evaluations or risk assessments have been included. Around 21.6\% of all service-ready models contained risk assessment headings in the README.md. Some integrated libraries, e.g. Autotrain\footnotemark[2], initialize default model cards based off the HF template. These often only contain text boilerplates for empty sections, such as "More information needed"\footnotemark[1]. For measurement integrity, we further filter out model cards with unfilled/auto-generated sections as non-compliant along that particular section. 

\footnotetext[1]{https://gitHub.com/huggingface/huggingface\_Hub/blob/main/src/huggingface\_Hub/templates/modelcard\_template.md}
\footnotetext[2]{https://huggingface.co/autotrain}
Evaluating ${CO_{2}}$ footprint is the other crucial sociotechnical component of model cards. It follows closely on the heels of the social impact of AI and concerns the broader impacts of the development cycle on the environment ~\cite{lacoste2019quantifying,strubell2020energy}. Conscientious, responsible developers may be motivated to report the social impact of the model comprehensively. Evaluation and disclosure of model ${CO_{2}}$ emissions is included as a dichotomous control. We use the API for headers and string matching to detect valid ${CO_{2}}$ emission entries under designated sections in the card text.

\textit{Competitive Benchmarking:}

The second research question tests for any significant association between model performance and documentation of risks and use limits. Over time, different leaderboards have been created to test different AI applications (see ~\ref{sec:respAI}). Leaderboard positions are highly vied, with high performers enjoying considerable visibility and popularity. The Open LLM Leaderboard is the most prominent and active leaderboard on Hugging Face, with its first edition running from May 2023 to June 2024. It was mainly geared towards language technologies and ranked submissions on aggregate performance across six extremely popular benchmarks~\cite{cobbe2021training,sakaguchi2021winogrande,clark2018think,zellers2019hellaswag,hendrycks2021measuring,lin2022truthfulqa}. With 7173 unique, complete submissions, the leaderboard has served to encourage performance validation while the community also consults it for model selection.  

The choice of the open LLM leaderboard was motivated by its prominence, upkeep, thoroughness, and monitoring. Notable leaderboard hosting services like Papers with Code~\cite{paperswithcode} are often static and primarily cover academic and research communities. HF teams continuously test submissions for authenticity and reproducibility. Unlike other community leaderboards like Kaggle, HF encourages users to proactively identify and report malpractice, such as contaminated models ~\cite{magar2022data}. This serves as a convenient sandbox for our research questions. 

We use leaderboard archives\footnotemark[3] to collect details on participating models. Submissions were ranked by their aggregate performance across six popular benchmarks. Details on the benchmarks used by the leaderboard and their metrics are provided in the appendix. Developers often submit multiple entries to report incremental increases in accuracy. While this ostensibly reflects innovation, resubmissions are often overfitted to perform ~\cite{hardt2017climbing, ethayarajh-jurafsky-2020-utility}  and may indicate the competitiveness of the participant rather than sustainable development. For our analysis, we only consider the best performance for each model, controlling for the number of attempts and the precision of the model version tested. We additionally control for evaluation malpractice through flagged models. 

Developers or platform moderators often assign 'not-for-all-audience'\footnotemark[4] and 'NSFW' tags to certain projects inappropriate for general use, such as ones trained on or meant for sexual content generation. By violating the fundamental premise of ethical AI, they are expected to show limited compliance. High-risk applications were incorporated as a categorical control in our analysis. 
\footnotetext[3]{https://huggingface.co/datasets/open-llm-leaderboard-old/results}
\footnotetext[4]{https://huggingface.co/content-guidelines}

\section{Results}
\subsection{Exploratory Analysis}
\label{sec:exploratory}
Before answering our quantitative RQs (covered in the next section), we begin with a preliminary analysis to understand general trends within our collected data.

\begin{figure}[!ht]
\centering
\frame{
\includegraphics[width=.95\textwidth,height=0.4\paperheight]{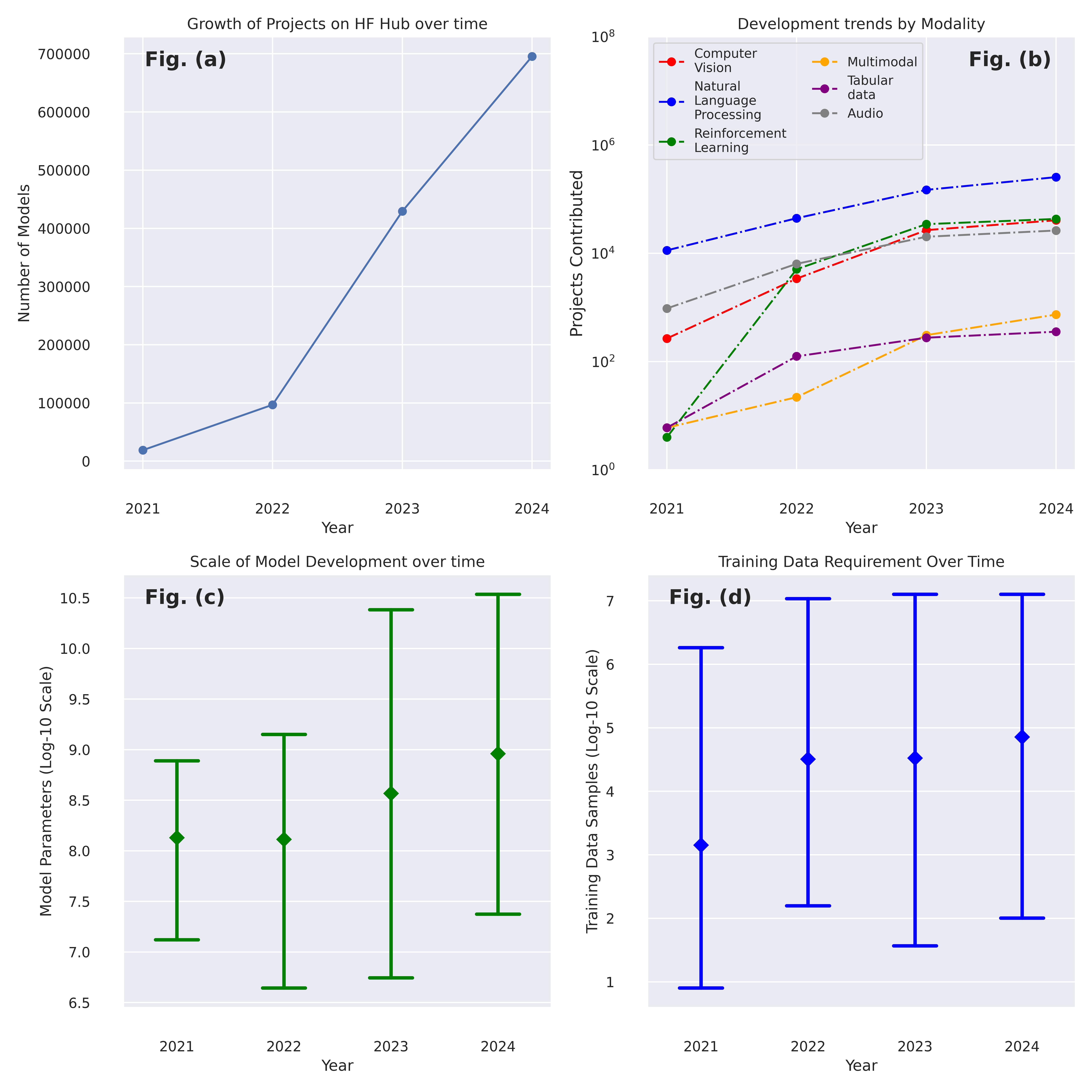}}
\caption{(a) Growth of the number of projects on HF Hub after the release of their client library in December 2020. (b) We also map development trends as the number of projects by modality among service-ready models uploaded since 2021. Natural Language Processing is consistently the most sought-after AI/ML application, closely followed by Reinforcement Learning, Computer Vision, and Audio. The trend over time (mean with 95\% confidence interval) in (c) model sizes and (d) training data requirements among 140,007 and 17,251 projects uploaded since 2021 showed a discernible increase in development scale.}
\label{fig:time_scale}
\end{figure}

Hugging Face initially gained recognition through an open-source implementation~\cite{wolf2020transformers} of the seminal transformer architecture~\cite{vaswani2017attention}, primarily targeted toward NLP development. The first version of the Hugging Face client library was released in late December 2020\footnotemark[5] to facilitate remote, collaborative development and artifact storage, reuse, and sharing. By the end of 2020, the collective comprised 4,634 projects and 672 unique contributors. The platform has since expanded support to over 20 ML libraries, AI frameworks, and applications. At the time of data collection on 06/15/2024, the HF Hub held 700,072 projects across 178,030 developers. Fig. ~\ref{fig:time_scale} charts the growth in repositories since the release of the Hub client.
\footnotetext[5]{https://pypi.org/project/huggingface-Hub/}

\begin{figure}[!htbp]
\centering
\frame{
\includegraphics[width=.95\textwidth,height=0.3\paperheight]{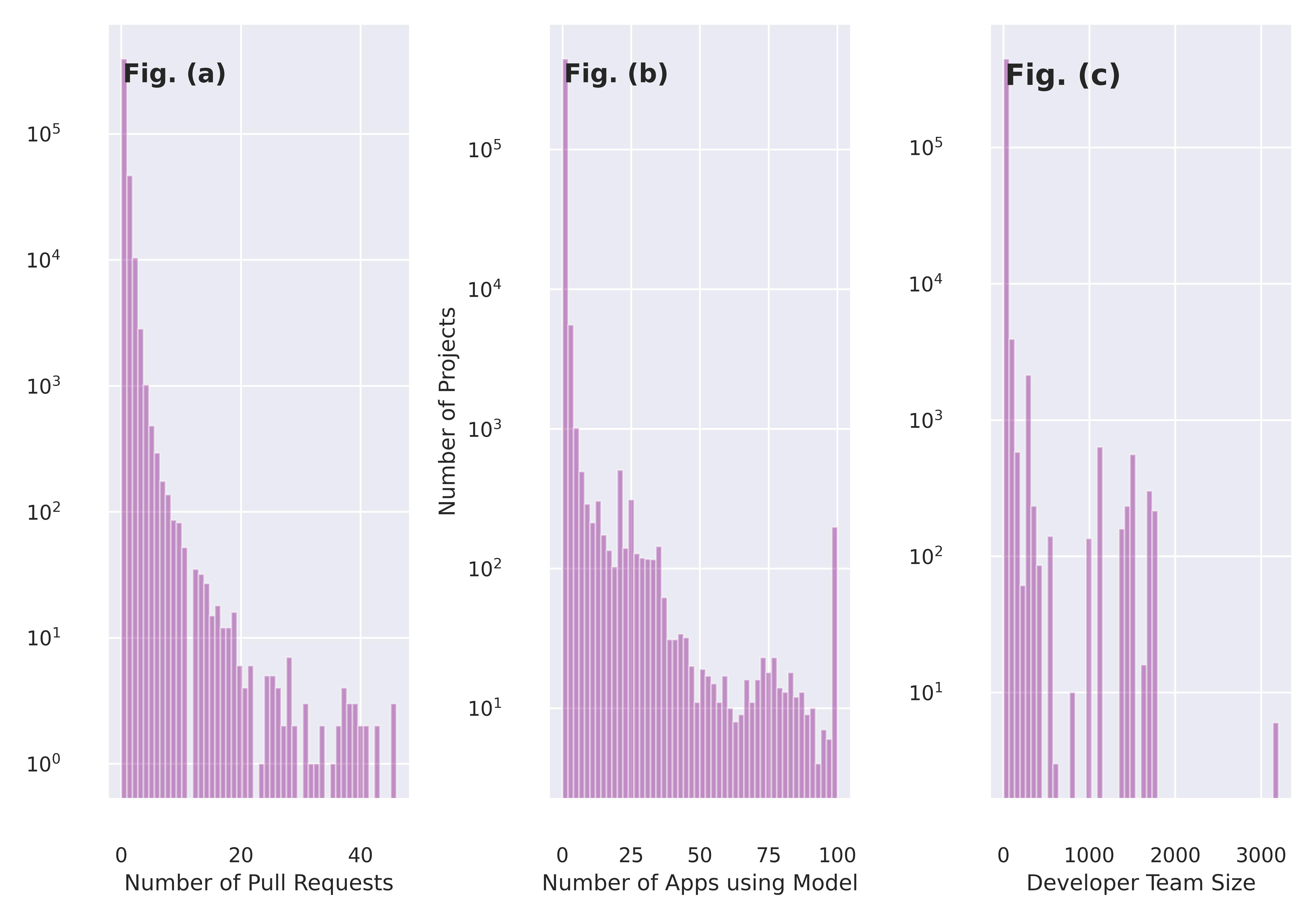}}
\caption{Contribution trends among all 456,545 service-ready projects. Most contributions are individual projects rather than collaborative, with 87.57\% having 0 pull requests and 86.34\% uploaded by single-member accounts. Around 5.46\% models have been deployed in applications (spaces).}
\label{fig:dev_trends}
\end{figure}

The democratization of innovation afforded by open sourcing, coupled with rapid progress in AI, has paved the way for training libraries and solutions to cater to all sizes and requirements. Hugging Face and its integrated libraries provide off-the-shelf options from industrial-scale foundation models to easy, low-compute customization. Besides team collaboration, the empowerment of individual developers was evident in a visual exploration of contribution patterns across the service-ready projects (see Fig. ~\ref{fig:dev_trends}). Around 87.57\% of all service-ready projects were contributed by individual accounts, and a staggering 86.34\% were built without receiving any collaborative input through pull requests. We also find that only 5.46\% of all projects see any downstream use in apps.

The Hub indexes most service-ready models by semantic categories that are system-generated or developer-annotated. Examining models by modality and application category tags, we find Natural Language Processing still accounts for most applications, followed by Reinforcement Learning, Computer Vision, and Audio applications. The release of large GPTs~\cite{achiam2023gpt,brown2020language} and diffusion models~\cite{rombach2022high} marked watershed moments for both language and image technologies. They were soon followed by a remarkable drive to build viable, open-access versions of commercial AI solutions. The dominance of NLP may also stem from Hugging Face's historic focus on NLP and more mature platform support. Computer Vision applications exceeded Audio projects by 2023. Reinforcement Learning sees a steady uptick between 2021-2023, parallel to the expansion of feedback-driven learning from simulating games to refining language applications~\cite{ouyang2022training}. All in all, these visualizations confirm general trends in AI and merit future exploration into the transfusion of innovation milestones in wider open-source practice.

We obtained verifiable model sizes (in parameters) and training data samples for 140,783 and 17,260 of the service-ready projects, respectively. Examining the temporal evolution of model sizes and training data requirements among post-2020 uploads in these subsets (Fig. ~\ref{fig:time_scale}), we observe a clear upward trajectory in development scale, favoring more sophisticated and data-intensive models.

\begin{figure}[!ht]
\centering
\frame{
\includegraphics[width=\textwidth,height=0.3\paperheight]{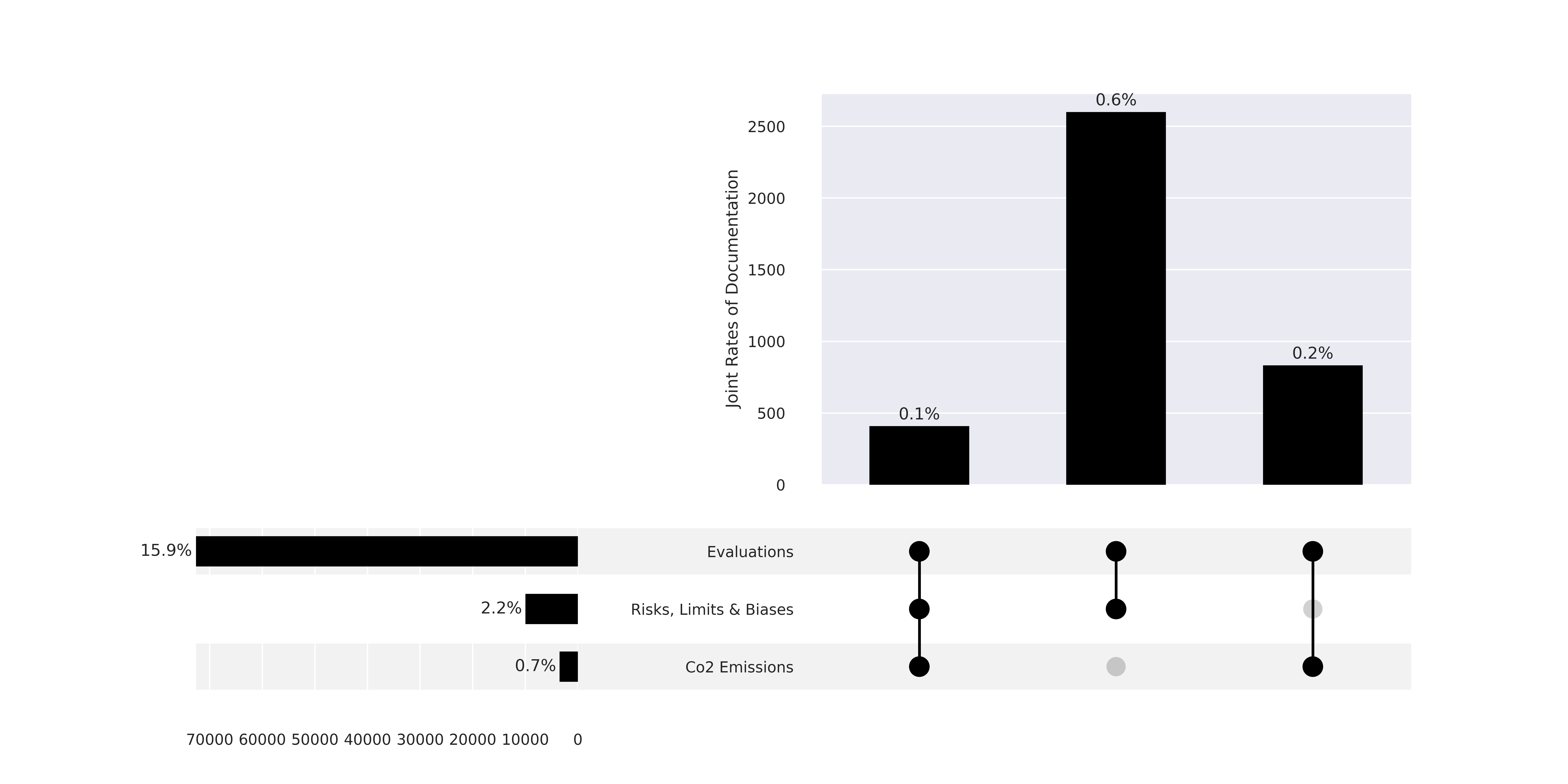}}
\caption{Documentation Rates between of Model Card components. Among 456,545 usable models, evaluations were most documented (15.9\%), while risks and limitations were found among 2.2\%. Finally, ${CO_{2}}$ emissions saw the least reporting at 0.7\%. About 0.7\% contained both evaluations and limitations. Only around 0.1\% of the models complete all three sections.}
\label{joint_compliance}
\end{figure}

An overall analysis of model card reporting among 456,545 service-ready models found generally low model card compliance and even differences between different sections (see Fig. ~\ref{joint_compliance}). Evaluations were most documented (15.9\%), while risks and limitations were found among 2.2\%. Finally, ${CO_{2}}$ emissions saw the least reporting at 0.7\%. About 0.7\% contained both evaluations and limitations. Only around 0.1\% of the models complete all three sections. These findings broadly agree with trends seen in prior work on AI documentation ~\cite{liang_whats_2024,yang_navigating_2024} and call for greater attention to documentation and comprehensiveness across all reporting requirements.

\begin{figure}[!ht]
\centering
\frame{
\includegraphics[width=.95\textwidth,height=0.25\paperheight]{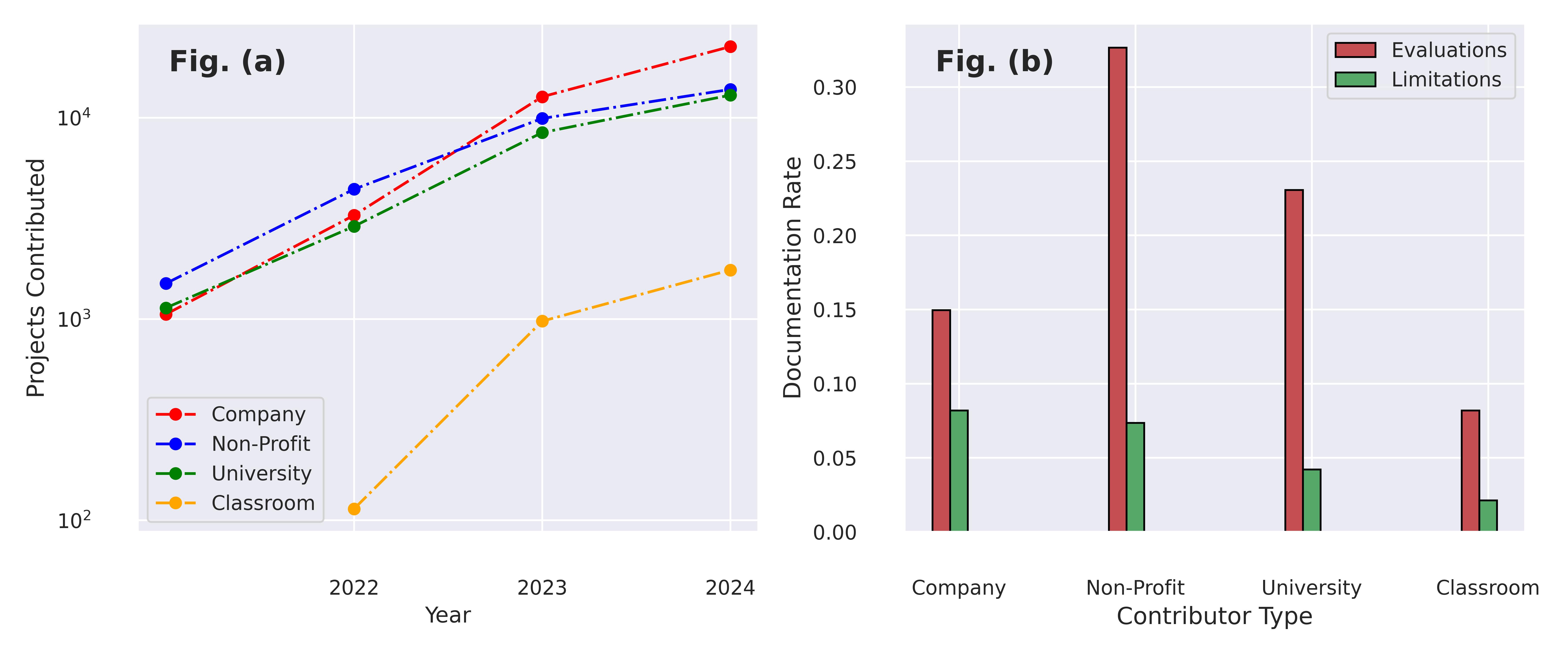}}
\caption{a. We find company contributions showing the highest growth rate between 2022 and 2024., surpassing universities and non-profits b. Noticeable differences exist in documentation across different developer/provider types. Non-profits lead among organizations in providing model evaluations. On the whole, non-profits, companies, and universities document risks more than the population average.}
\label{fig:entity_growth_compliance}
\end{figure}

Analyzing projects by team ('organization') types in Fig. ~\ref{fig:entity_growth_compliance} a. we find company contributions showing the highest growth rate between 2022 and 2024. By 2024, company created models exceed traditional OSS participants such as academia and non-profits. Meanwhile Fig ~\ref{fig:entity_growth_compliance} b. shows noticeable differences in documentation across different developer/provider types. Non-profits lead among organizations in model evaluation and documentation. Yet, on the whole, non-profits, companies, and universities document risks more than the population average.

\subsection{Multivariate Hypothesis Testing for RQ1 and RQ2}
\label{sec:empirical}
Here, we address our core RQs introduced above. For both RQs, we consider models whose information on significant covariates was released on the HF Hub and available in structured, parseable form. Based on the literature review, trends in AI safety research, and our exploratory analysis, we consider five main categories of covariates: project scale, modality, domain, popularity and usage, and developer engagement. This left us with 7093 samples for RQ1, a subset of the entire HF directory representing service-ready, highly transparent models where data provenance and model specifics (number of parameters) are available through safe, robust file management. Around 23.19\%, 7.86\% and 2.04\% had provided evaluation, risk assessments, and ${CO_{2}}$ emission data, respectively.            
We frame our RQs as binary prediction modeling to determine if risk assessment and social impact accountability are significantly associated with 1. rates of performance evaluation and disclosure and 2. absolute mean performance on a set of very popular benchmarks used in the Open LLM Leaderboard. For both cases, we model the likelihood of risk assessment in model cards using binomial logit models, where evaluation practices (RQ1) or performance (RQ2) are the main regressors of interest, adjusting for crucial project-level covariates. We set the significance level of our analysis at 0.01. 
 
RQ2 is based on a subset of RQ1, which participates in the open LLM leaderboard and only comprises NLP models. In this particular subset of models with 100\% evaluation reporting (through benchmark participation), we again find higher than average risk (8.7\%) and carbon emission (1.5\%) reporting, around 7173 models completed all six benchmarks. We test RQ2 on a smaller subset of 789 of these models for which all significant covariates were available. Some generalist models spanned multiple domains, such as medicine and legal/financial applications, leading to the aliasing. These knowledge domains were merged into a single category and renamed “multi-domain.”

    Numeric covariates were log-transformed (base 10) for skew correction and comparison along the scale of different projects, followed by standardization. We check for multicollinearity using the \textit{car} package from R, removing variables with VIF factor > 5. This excluded the model domains 'music' and 'art' from RQ1 and 'Biology,' 'Chemistry' and precision category 'Torch Float16' from RQ2.  We checked for high-leverage outliers for RQ1, based on Cook’s distance (${D > 4/N}$) and standard residuals ($> 3$), and removed 1 data point each from both analyses. Using the Box-Tidwell approach~\cite{box1962transformation}, suitable higher-order transformations (See Table. ~\ref{tab:rq1} And Table. ~\ref{tab:rq2}) were performed on some variables to ensure that assumptions of linearity between log odds and the predictors held. Compared to simpler, more interpretable models, the models with power-transformed variables were ultimately preferred for the final reporting due to greater explainability (AIC from 374.4 to 371.6 for RQ2 and from 3454.3 to 3410.9) and validity. Encouragingly, the significant effects and their directionality remain largely preserved across both approaches, confirming the robustness of the results. All effects significant in the transformed models are also significant in the simpler models, except for the number of models built by the developer and likes in RQ1, which do not appear significant before transformation.

    Finally, residual tests were performed using the \textit{DHARMa} package. Neither regression model showed significant dispersion (RQ1:${p=0.82}$ and RQ2:${p=0.85}$), presence of outliers (RQ1:${p=0.18}$ and RQ2:${p=0.42}$), or deviation from normal distributions (KS test; RQ1:${p=0.78}$ and RQ2:${p=0.78}$).


\begin{table}[!ht]
\begin{tabularx}{\textwidth}{|X|X|X|X|}
\hline

 & \textbf{Predictor} & {\hfil \textbf{Coefficient}}& {\hfil\textbf{p-value}}  \\ \cline{1-4}
 & (Intercept) & -3.263988 &  \textbf{$<$0.0001}  \\  \hline
 
\multirow{2}{*}{\textbf{Model Scale}} 

& Parameters \footnotemark[1] & 0.120302 & 0.026334  \\ \cline{2-4}
& Data size \footnotemark[1] & 0.179451 & \textbf{0.000169} \\ \hline

\multirow{2}{*}{\textbf{Modality}}  

& Audio & -0.949783 & \textbf{0.003028} \\ \cline{2-4}
& Computer Vision & 0.815687 & 0.014426  \\ \cline{2-4}
& Multimodal & -1.288032 & 0.209746 \\ \cline{2-4}
& Natural Language Processing & 0.384321 & 0.019977  \\ \cline{2-4}
& Reinforcement Learning & -13.517337 & 0.984074 \\ \hline

\multirow{2}{*}{\textbf{Domain}}                             
& Biology & 0.068115 & 0.914340 \\ \cline{2-4} 
& Chemistry & 0.320644 & 0.713684 \\ \cline{2-4}
& Climate & 1.646929 & 0.324356 \\ \cline{2-4}
& Code & -0.483826 & 0.135364 \\ \cline{2-4}
& Finance & -0.174213 & 0.796490 \\ \cline{2-4}
& Legal & 0.191019 & 0.720080 \\ \cline{2-4}
& Medical & -1.235878 & 0.035409 \\ \hline

\multirow{2}{*}{\textbf{Model Developer}}
& Team members \footnotemark[1] & 0.193956 & \textbf{0.000149} \\ \cline{2-4}
& Total models \footnotemark[2] & 0.248375 & \textbf{$<$ 0.0001} \\ \cline{2-4} 

& Company & -0.204922 & 0.273141 \\ \cline{2-4}
& University & -0.005852 & 0.983995 \\ \cline{2-4}
& Classroom & 0.623096 & 0.445068 \\ \cline{2-4}
& Non-profit & 0.529477 & 0.021068 \\ \hline

\multirow{2}{*}{\textbf{Use and Popularity}}

& Likes \footnotemark[3] & 0.174444 & \textbf{0.001643} \\ \cline{2-4}
& Number of Spaces \footnotemark[1] & -0.015841 & 0.713853 \\ \hline

\multirow{2}{*}{\textbf{Repository Activity}}

& Total Commits \footnotemark[2] & -0.359598 & \textbf{$<$ 0.0001} \\ \cline{2-4}
& Threads \footnotemark[1] & 0.090050 & 0.026619  \\ \cline{2-4}
& PR \footnotemark[1] & 0.001501 & 0.974375 \\ \cline{2-4}

& Repository age \footnotemark[2] & 0.054596 & 0.268635 \\ \hline

\multirow{2}{*}{\textbf{Transparency}} 

& ${CO_{2}}$ footprint & 2.177332 & \textbf{$<$ 0.0001} \\ \cline{2-4}
& Evaluation Availability & 0.913310 & \textbf{$<$ 0.0001} \\ \hline

\multirow{1}{*}{\textbf{Others}} & High Risk Application & -13.834954 & 0.968635 \\ \hline

\multicolumn{2}{|c}{}  & \multicolumn{2}{|c|}{\textbf{N= 7092 ${R^{2}}$ = 0.115}} \\ 
         
        \multicolumn{2}{|c}{}  & \multicolumn{2}{|c|}{\textbf{ AIC = 3411 }} \\ \hline

         
\end{tabularx}

\begin{flushleft}
\footnotesize $^1$ Log transformed (base 10) and Standardized $^2$ Log (base 10), ${1/x}$ and Standardized $^3$ Log (base 10), $x^{0.3}$ and Standardized
\end{flushleft}

\caption{Test statistics for binomial logistic regression of limits, bias, and risks documentation rates among models based on 1. their project attributes, 2. rates of compliance with related components of the Model Card}

\label{tab:rq1}
\end{table}



\begin{table}[!ht]
\begin{tabularx}{\textwidth}{|X|X|X|X|}
\hline
 & \textbf{Predictor} & {\hfil \textbf{Coefficient}}& {\hfil\textbf{p-value}}  \\ \cline{1-4}
 
& (Intercept) & -2.8854 & \textbf{$<$ 0.0001} \\\hline

\multirow{2}{*}{\textbf{Model Scale}} 

& Parameters \footnotemark[2] & 0.6695 & \textbf{0.000803} \\ \cline{2-4}
& Data size \footnotemark[1] & -0.1617 & 0.371708 \\\hline

\multirow{2}{*}{\textbf{Domain}}                             
       
& Multi-domain & 17.3876 & 0.987295 \\ \cline{2-4}
& code & -16.6237 & 0.987853 \\ \cline{2-4}
& medical & 0.5741 & 0.730284 \\  \hline

\multirow{5}{*}{\textbf{Model Developer}}

& Team members \footnotemark[1] & 1.0562 & \textbf{$<$ 0.0001} \\ \cline{2-4}
& Profile models \footnotemark[1] & -0.6927 & \textbf{0.000257} \\ \cline{2-4}
& Company & -1.6773 & \textbf{0.003041} \\  \cline{2-4}
& University & -0.2654 & 0.714144 \\ \cline{2-4}
& Non profit & -1.3751 & 0.173400 \\ \hline

\multirow{2}{*}{\textbf{Use and Popularity}}

& Likes \footnotemark[1] & -0.3491 & 0.194646 \\ \cline{2-4}
& Number of Spaces \footnotemark[1] & 0.2701 & 0.155802 \\ \hline

\multirow{3}{*}{\textbf{Repository Activity}} 
& Total Commits \footnotemark[1] & 0.8053 & \textbf{$<$ 0.0001} \\ \cline{2-4}
& Threads \footnotemark[1] & 0.1761 & 0.427108 \\ \cline{2-4}
& PR \footnotemark[1] & -0.1766 & 0.162182 \\  \cline{2-4}
& Repository age \footnotemark[1] & -0.0203 & 0.920262 \\  \hline

\multirow{1}{*}{\textbf{Transparency}} & ${CO_2}$ Footprint availability & 2.3698 & \textbf{0.001487}  \\ \hline
 \multirow{1}{*}{\textbf{Evaluation Details}}   
 
 & Accuracy \footnotemark[3] & -0.7631 & \textbf{0.001124} \\ \cline{2-4}
& Flagged & -0.2596 & 0.796655 \\ \cline{2-4}
& Attempts \footnotemark[1] & 0.3128 & 0.038215  \\ \hline

 \multirow{2}{*}{\textbf{Precision}} 
 
& 4 bit & -18.0137 & 0.993056 \\ \cline{2-4}
& 8 bit & -0.3797 & 0.802545 \\ \cline{2-4}
& Torch BFloat16 & 0.3294 & 0.316380 \\ \hline

\multirow{1}{*}{\textbf{Others}} & High Risk Application & -15.3855 & 0.994607\\ \hline

     \multicolumn{2}{|c}{}  & \multicolumn{2}{|c|}{\textbf{N= 788 ${R^{2}}$ = 0.272}} \\ 
         
        \multicolumn{2}{|c}{}  & \multicolumn{2}{|c|}{\textbf{ AIC = 371.636 }} \\ \hline
\end{tabularx}
\begin{flushleft}
\footnotesize $^1$ Log transformed (base 10) and Standardized $^2$ Log (base 10), ${x^{4.5}}$ and Standardized $^3$ Standardized
\end{flushleft}
\caption{Test statistics for binomial logistic regression of limits, bias, and risks documentation rates among models based on 1. features of leaderboard models 2. competitive performance of the models}
\label{tab:rq2}
\end{table}

\textbf{Interpretation}
Our analysis from RQ1 confirms a strong association between evaluation practices and risk documentation, with models reporting some form of evaluation being 149.2\% more likely to also carry information on model risks and limits. Other positive effects come from training data size, documentation of ${CO_{2}}$ footprint, developer team size, commit activity, and popularity (likes). Audio applications and models associated with high contributors (more models) are also less likely to carry risk documentation. 

Meanwhile, RQ2 finds that high performers on the Open LLM Leaderboard are less likely to document risks and limitations. One standard unit increase in accuracy reduced risk reporting chances by 53.4\%. Greater model size (parameters), documentation of ${CO_{2}}$ footprint, high number of commits, and developer team size also predict higher chances of a project carrying such documentation. At the same time, companies and high contributors are less likely to do the same. Interestingly, specific model knowledge domains do not exert any significant effect across both analyses, i.e., risk reporting rates are relatively the same across high-stake applications such as medicine or finance, niches such as code, and all other general domains. 

\section{Discussion}

Evaluation is core to responsible AI. It is essential to determining model capabilities, and also serves as empirical means to other aspects of responsible AI, such as understanding and acknowledging risks and limitations. Our analyses of OSS practitioners confirms that evaluation and risk assessment generally go hand in hand. However, we also observed that metric-centric arenas, such as competitive leaderboards, may see lesser acknowledgment of model risks among high performers. 

Certain other observations were consistent across both analyses. As one might anticipate, development at scale (data-intensive training or parameterization) positively correlates with compliance. Documentation of social impact is also closely associated with broader awareness (as expressed through estimation and reporting of ${CO_{2}}$ footprint), and projects with more activity, contributions and larger teams tended to do a better job with risk reporting. On the other hand, prolific developers appear to pay less attention to assessing and documenting the limitations of their projects. Informed by these trends, we hereby present our recommendations for contributors, entrepreneurs, and AI hosting services. These include practices and interventions to encourage documentation overall, and to improve efficacy of evaluation protocols in informing both model strengths and weaknesses. 

\subsection{Recommendations} 

As one of the leading open-source AI hosting services, Hugging Face has taken steps to inculcate responsible documentation, monitor compliance\footnotemark[1], and keep up with regulation~\cite{huggingfacePublicPolicy}. Results from our empirical analyses suggest that risk documentation practices are more prevalent among large teams, while most contributions come from individual developers. Model card guidelines used by Hugging Face and other notable institutions\footnotemark[2]\footnotemark[3] are detailed to facilitate auditing, and usually set specific tasks across developers, sociotechnical experts and managers. Risk assessments involve multiple roles and can make compliance overwhelming for small teams. Streamlining, such as outlining priority requirements may make risk documentation more approachable. 

HF's open LLM leaderboard is a massive undertaking, supported by collaborative monitoring labor from community and moderators. It is notably more transparent than conventional leaderboards (See Sec. ~\ref{sec:respAI}) and tracks model size, precision, libraries, and architectures of most submissions. Such considerations are expected to support explainability, promote sustainable models and inform judicious model selection for small-scale, decentralized applications, which are often less resourced than larger communities or funded corporations. 
 
Data providers and platforms hosting leaderboards need to consider the emerging needs of evaluation, improve upon the reported limitations of benchmarking, and consider multi-faceted tasks and metrics -- in short, make evaluation more multi-dimensional. Beyond leaderboards, the choice of tests and metrics for all other model evaluations (and risk assessment) are generally left up to the developer's discretion. While HF modelcards mention that evaluation choices should ideally address social impact, there is currently a gap in terms of standards, expectations, and norms.  Our results suggest that more precise guidelines could have a very large impact on OSS reporting practices. Lastly, fostering broad awareness and hosting training modules on the different dimensions of AI risks (social and environmental), promoting well-documented models~\cite{liang_whats_2024}, and messaging on the importance of quality and safety of models (over quantity) can improve overall developer accountability.

 \footnotetext[1]{https://huggingface.co/spaces/society-ethics/model-card-regulatory-check/}
 \footnotetext[2]{https://bias.xd.gov/resources/model-card-generator/}
 \footnotetext[3]{https://huggingface.co/docs/Hub/en/model-card-appendix}

\subsection{Limitations}

Our research formulation, analyses, and results are meant to explore the correlation between evaluation and risk documentation rather than establish a causal implication. We aimed to measure the prevalence of responsible development practices as operationalized through reporting compliance. With limited regulatory requirements or platform specifications on OSS as of now, we cannot conclusively determine whether the risk assessments provided are necessary or sufficient for any given model, i.e. while our quantitative analyses help to explain existing behaviors, it can be hard to translate these behaviors into impact. Growing consensus over AI safety standards and research establishing testing protocols can be expected to be adopted by platforms, and future work may explore their diffusion into practice, particularly how specific tests enable risk quantification and their efficacy.

Hugging Face's popularity and moderation make their leaderboards amenable for our research questions (See Sec. ~\ref{sec:rq} and Sec. ~\ref{sec:data}). Most leaderboards are built towards a particular domain and set of tasks, and submissions are generally uniform in modality. Despite the testing ground being an NLP-only leaderboard, our conclusions about developer accountability are expected to hold for rapidly growing technologies, invariant of modality. We explored popular\footnotemark[6] HF leaderboards like MTEB\footnotemark[4] and LMSys\footnotemark[5] for additional scenarios and found that they drew much lower participation, with too few fully open projects for representative quality or power. We look forward to future studies on improved, up-and-coming leaderboards to further validate our findings and inform evaluation practices going forward.  

 \footnotetext[4]{https://huggingface.co/spaces/mteb/leaderboard}
 \footnotetext[5]{https://huggingface.co/spaces/lmsys/chatbot-arena-leaderboard}
\footnotetext[6]{https://huggingface.co/docs/leaderboards/en/leaderboards/intro}

\subsection{Ethics Statement}
For our data collection, we largely followed prior work and used the public Hugging Face API for model cards and open repository data. Beyond the API, we collected some limited public-facing numeric data from model landing pages, which are intended for public reference and sharing with no expectation of privacy (Table. ~\ref{tab:apps}). We did not add any features to our data collection code for specifically parsing  personally-identifying information, nor was such information required for our analysis. The only identifiers used were public developer usernames, which are also part of the model path in the HF web indexing. 
Finally, we note that some public, open weight models carry ''Not Safe For Work" and ''Not for all Audiences" warnings from developers or the platform moderators\footnotemark[7]. We retain these labels in our dataset so that any researchers wishing to use this data can be fully informed about the potential for some model metadata to contain content inappropriate for some settings.
Proprietary LLM-based language editing services Grammarly and Anthropic Claude were used to a limited extent to correct misspellings, grammar and consistency of composition, and the resulting manuscript was thoroughly verified and updated by all the authors over multiple iterations. 

\footnotetext[7]{https://huggingface.co/content-guidelines}
\section{Conclusion}
Through our focused study of a rising open source platform, we had the opportunity to observe a diverse range of AI/ML applications and development practices. Our analyses empirically probe open source AI trends in the backdrop of increasing concerns over their potential to transform or affect society, and consequent legal and ethical oversight. As we situate our investigation amidst the interests of these various stakeholders, we discover promising trends of concurrent compliance of evaluations and risk assessments. At the same time, our large sample study produces evidence supporting long standing observations and calls for fundamental reforms and greater rigor in AI evaluation. 

As AI continues to grow, these lessons emphasize the importance of fostering a culture of responsible development and accountability across all sectors, not only commercial but also informal and non-profit undertakings. Platforms, developers, and stakeholders must work together to establish best practices and design balanced policies and standards that mutually support each other while also preserving the true spirit of innovation. This will be vital in ensuring that AI technologies are developed and deployed ethically, safely, and benefit humanity at large.

\bibliographystyle{plain} 
\bibliography{sample-base.bib}
\appendix

\section{Benchmarks in Open LLM Leaderboard}
\begin{table}[h]
\centering
\begin{tabular}{|p{0.3\textwidth}|p{0.4\textwidth}|p{0.2\textwidth}|}
\hline
\textbf{Benchmark} & \textbf{Brief Description} & \textbf{Metric Used} \\
\hline
AI2 Reasoning Challenge (ARC)~\cite{clark2018think} & A set of grade-school science questions (25-shot) & Accuracy (normalized by target length)  \\
\hline
HellaSwag~\cite{zellers2019hellaswag} & A test of commonsense inference, challenging for SOTA models but easy for humans (10-shot)& Accuracy (normalized by target length) \\
\hline
MMLU~\cite{hendrycks2021measuring} & Measures multitask accuracy across 57 tasks including mathematics, history, law, and more (5-shot) & Accuracy \\
\hline
TruthfulQA~\cite{lin2022truthfulqa} & Measures a model's propensity to reproduce common online falsehoods (0-shot) & MC2 (Normalized probability over true references)  \\
\hline
Winogrande~\cite{sakaguchi2021winogrande} & An adversarial and difficult Winograd benchmark for commonsense reasoning (5-shot) & Accuracy  \\
\hline
GSM8k~\cite{cobbe2021training} & Diverse grade school math word problems to test multi-step mathematical reasoning (5-shot) & Accuracy  \\
\hline
\end{tabular}
\caption{Summary of the six key benchmarks adopted by the Open LLM Leaderboard v1 from the Eleuther LLM evaluation harness~\cite{eval-harness}. The main leaderboard, by default, ranks models by their average performance across these benchmarks}
\label{tab:open-llm-benchmarks}
\end{table}

\section{Service-ready Features and Identifiers}

\begin{itemize}
    
    \item \textbf{Use this code}: Platform-generated example scripts (button above the repository) to guide model loading and use through recognized libraries.
    
    \item \textbf{Endpoints\_compatible\footnotemark[1]}: Inference Endpoints are scalable and production-ready API endpoints for machine learning models. This repo tag indicates that a particular model is compatible with Inference Endpoints. 
    
    \item \textbf{Pipeline\_tag\footnotemark[2]}: These tags denote the specific task a model was designed for, such as "text-classification", or "object-detection". These tags act as semantic categories to enhance model discoverability for specific applications, and are either detected by the hub or indicated by the developer from a list of recognized applications.
    
    \item \textbf{Autotrain\_compatible\footnotemark[3]}: This tag indicates if a project is a complete pre-trained model and compatible within the HF ecosystem for downstream fine-tuning on custom data.
    
    \item \textbf{Text-embeddings-inference\footnotemark[4]}: Allows generation of text embeddings at scale from compatible models. This tag appears in the repository of such compatible models.
    
    \item \textbf{Text-generation-inference\footnotemark[5]}: A runtime, sometimes also a widget, to handle text generation queries to a model. Enabled for fully functional LLMs, these models are tagged the same. 
    
    \item \textbf{Intended Purpose (Documentation)\footnotemark[6]}: We use the model card scanner from HF to detect detailed documentation from developers on explaining model usage, e.g., direct or downstream applications, optionally provided code etc
\end{itemize}

\footnotetext[1]{https://huggingface.co/inference-endpoints/dedicated}
\footnotetext[2]{https://huggingface.co/docs/hub/en/models-tasks}
\footnotetext[3]{https://huggingface.co/autotrain}
\footnotetext[4]{https://github.com/huggingface/text-embeddings-inference}
\footnotetext[5]{https://github.com/huggingface/text-generation-inference}
\footnotetext[6]{https://huggingface.co/docs/hub/en/model-card-annotated}

\end{document}